# Парадокс друга Вигнера: объективной реальности не существует?

А.В.Белинский

Физический факультет МГУ им. М.В.Ломоносова

Показано, что отсутствие объективного существования результатов квантовых измерений не может быть доказано известными экспериментами. Приведены также аргументы общего характера, подтверждающие это заключение.

### 1. Введение

В последнее время заметно возрос интерес к выяснению онтологического статуса волновой функции и вектора квантового состояния, являющимися одними из основных понятий квантовой механики. Проявления так называемой квантовой нелокальности и множество квантовых парадоксов не находят бесспорных и общепринятых непротиворечивых интерпретаций. В этой связи все больше приверженцев находит так называемая информационная интерпретация, истоки которой намечены еще Нильсом Бором [1], получили дальнейшее развитие, например, в работах [2,3].

Информационная интерпретация отрицает объективное существование волновой функции и вектора состояния, приписывая им статус математических абстракций, исключительная роль которых сводится лишь к инструменту вычислений. Это сразу снимает ряд вопросов к большинству квантовых парадоксов, например, нет проблем с мгновенным коллапсом волновой функции, т.к. ее в природе не существует, с нелокальностью, поскольку все ее проявления опять же связаны с необычным поведением вектора состояния, и т.д.

Одним из аргументов в пользу такой интерпретации выдвигается так называемый парадокс друга Вигнера, получивший, по утверждению авторов [4], экспериментальное подтверждение.

Кратко, сущность парадокса сводится к следующему. Друг Вигнера производит измерение квантовой системы, находящейся в состоянии квантовой суперпозиции. В результате этого измерения вектор состояния коллапсирует и получается определенный результат. Но сам Вигнер о нем не знает. Для него, поэтому, система по-прежнему находится в состоянии суперпозиции. Что же произошло на самом деле? Был коллапс, или нет? Если друзья не общаются между собой, то у каждого получается своя "объективная" реальность. Конечно, ситуация здесь преднамеренно сильно упрощена, но такое вступление имеет целью подготовить читателя к анализу довольно сложного эксперимента [4].

### 2. Эксперимент, воспроизводящий парадокс друга Вигнера [4]

Пара запутанных по поляризации фотонов поступает к друзьям Алисы и Боба, каждый к своему в отдельную лабораторию. Сами же Алиса и Боб, также находящиеся в разных местах, имеют возможность невозмущающим образом либо получить тот же результат – значения Ao и Bo дихотомных переменных, равных +1 или −1 в зависимости от состояния поляризации зарегистрированных фотонов, либо осуществить, по утверждению авторов, невозмущающее измерение того, произошел ли коллапс состояния суперпозиции запутанных фотонов. Для этого у Алисы, как и у и Боба, приготавливается еще по паре запутанных фотонов для наблюдения двухфотонной интерференции. И на тех же детекторах при небольшой модернизации экспериментальной установки они получают также дихотомные значения $A_1$ и $B_1$, равные +1 или −1.

Итак, в каждом акте измерений существуют вполне определенные значения $A_o$ и $B_o$, т.е. объективно коллапс осуществился. Но Алиса и Боб видят при этом квантовую интерференцию, которая, якобы, свидетельствует об обратном. Как в этом предлагают убедиться авторы [4]? Они из величин $A_i$ и $B_i$ составляют неравенство Белла типа Клаузера-Хорна-Шимони-Хольта (CHSH) [5]:

$$S = \langle A_1 B_1 \rangle + \langle A_1 B_0 \rangle + \langle A_0 B_1 \rangle − \langle A_0 B_0 \rangle \leq 2, \qquad (1)$$

и в эксперименте оно нарушается, что свидетельствует об отсутствии определенных значений этих величин, хотя $A_o$ и $B_o$ известны. Все ли здесь корректно? Вряд ли нужно подчеркивать всю принципиальную значимость данного вопроса.

### 3. Особенности неравенства CHSH

Для выяснения следствий нарушения неравенства CHSH обратимся к простейшему его выводу [6,7]. Пусть каждая из четырех величин $A_i$, $B_i$ априорно имеет вполне определенные значения $a_o$, $a_1$, $b_o$, $b_1$, равные +1 или −1. Тогда из них можно составить следующие выражения:

$$a_1 b_1 + a_1 b_o + a_o b_1 − a_o b_o = a_1(b_1+b_o)+a_o(b_1−b_o) = b_1(a_1+a_o)+ b_o(a_1−a_o) = \pm 2,$$

откуда следует (1).

Но значит ли это, что (1) может нарушаться только при отсутствии *всех* определенных априорных значений $A_i$, $B_i$? Отнюдь нет. Достаточно лишь двух, например, $A_1$ и $B_1$, а $A_o$ и $B_o$ могут быть полностью определены. Неравенство (1) при этом может тоже нарушаться, как следует из (2), поскольку обе скобки могут быть ненулевыми, точнее неопределенными.

Другое доказательство (1) использует элементарные четырехмерные вероятности $P(A_o,A_1,B_o,B_1)$ [8]. Оно также строится на априорном существовании *всех* элементарных вероятностей $P(A_o,A_1,B_o,B_1)$. Но для нарушения (1) достаточно отсутствие существования *лишь некоторых*.

Итак, нарушение (1) вовсе не свидетельствует об отсутствии объективно реально существующих $A_o$ и $B_o$ и факта коллапса исходного вектора квантового состояния.

### 4. Некоторые общие соображения

Уже сам факт возможности невозмущающего измерения присутствия или отсутствия коллапса вектора состояния в удаленной локализованной системе вызывает ряд вряд ли разрешимых вопросов. Если коллапс происходит мгновенно (а этому существует и экспериментальное подтверждение, по крайней мере, скорость коллапса в [9,10] превысила *c* на несколько порядков), то имея возможность такого измерения я могу моментально передавать информацию сверхсветовым телеграфом, поскольку присутствие и отсутствие коллапса закодирую дихотомными значениями, соответствующими 1 биту. Но этому препятствует "no communication theorem" [11], имеющая весьма общий характер, так что преодолеть ее, как представляется, невозможно.

Действительно, предположим, что в эксперименте [1] Алиса и Боб производят интерференционный эксперимент *до* регистрации их друзьями запутанной пары фотонов, т.е. до коллапса. Естественно, они получат интерференцию, которая подтверждает отсутствие коллапса. Но что, если коллапс происходит *до* измерения Алисы и Боба? В полном соответствии с "no communication theorem" *ничего* не должно измениться, иначе у них с друзьями установится мгновенный сверхсветовой канал связи.

Итак, даже не вникая в тонкости эксперимента и особенности неравенства Белла типа CHSH, можно заключить, что отрицание существования объективной реальности не может быть доказано на основании парадокса друзей Вигнера.

**5. Заключение**

Какой вывод можно сделать из приведённой аргументации? Доказывает ли она несостоятельность информационной интерпретации квантовой механики? Вовсе нет. Но если бы удалось доказать отсутствие объективной реальности применительно к волновой функции и вектору состояния, то все остальные интерпретации следовало бы отправить в архив. Однако, как следует из всего вышеизложенного, это было бы преждевременно. Информационная интерпретация остаётся лишь одним из претендентов наряду с другими непротиворечивыми концепциями.

**Wigner's friend paradox: objective reality does not exist?**
**Alexander V. Belinsky**
*Department of mathematical modeling and informatics, Department of Physics of the Earth, Faculty of Physics, Lomonosov Moscow State University. Moscow 119991, Russia*

It is shown that the absence of an objective existence of the results of quantum measurements cannot be proved by known experiments. There are also general arguments confirming this conclusion.